\documentclass[aps,
physrev,
superscriptaddress,
twocolumn,
citeautoscript]{revtex4-2}

\usepackage{float}
\usepackage{amsmath}
\usepackage{amssymb}
\usepackage{amsfonts}
\usepackage{dsfont}
\usepackage{placeins}
\usepackage{graphicx,graphics}
\usepackage{color}
\usepackage[autostyle]{csquotes}
\usepackage{changes}
\usepackage{lipsum}

\usepackage{physics}

\usepackage{svg}
\usepackage{hyperref}

\hypersetup{
	pdfborder = {0 0 0}
}
\usepackage{verbatim}

\usepackage{orcidlink}

\newcommand{\normord}[1]{\vcentcolon \, \mathrel{#1} \, \vcentcolon}
\providecommand{\vcentcolon}{\mathrel{\mathop{:}}}
\newcommand{\avg}[1]{\langle{#1}\rangle}

\newcommand{\ii}{\mathrm{i}}
\newcommand{\ee}{\mathrm{e}}

\newcommand{\proj}{\hat{P}}

\newcommand{\dens}{\hat{\rho}}

\newcommand{\ham}{\hat{H}}
\newcommand{\loss}{\hat{L}}
\newcommand{\occ}{\hat{n}}

\newcommand{\ann}{\hat{a}}
\newcommand{\adag}[1][]{\hat{a}^{\dagger #1}}

\newcommand{\bnn}{\hat{b}}
\newcommand{\bdag}[1][]{\hat{b}^{\dagger #1}}

\newcommand{\affDLR}[0]{German Aerospace Center (DLR), Institute of Quantum Technologies, 89081 Ulm, Germany}
\newcommand{\affUniUlm}[0]{Institute for Complex Quantum Systems and IQST, University of Ulm, 89069 Ulm, Germany}
\newcommand{\affSherbrooke}[0]{Institut Quantique, Universit\'e de Sherbrooke, Sherbrooke, Qu\'ebec J1K 2R1, Canada}

\begin{document}

\title{Stroboscopic detection of itinerant microwave photons}

\author{Hanna Zeller \orcidlink{0009-0000-8264-1431}}
\email{Contact author: hanna.zeller@uni-ulm.de}
\affiliation{\affUniUlm}

\author{Lukas Danner}
\affiliation{\affDLR}
\affiliation{\affUniUlm}

\author{Max Hofheinz}
\affiliation{\affSherbrooke}

\author{Ciprian Padurariu}
\affiliation{\affUniUlm}

\author{Joachim Ankerhold}
\affiliation{\affUniUlm}

\author{Bj\"orn Kubala}
\affiliation{\affDLR}
\affiliation{\affUniUlm}

\date{\today}
\begin{abstract}
We present a novel scheme to detect  itinerant microwave radiation at the single photon level. Using existing Josephson-photonics devices, where two microwave cavities are coupled by a dc-voltage biased superconducting junction, we theoretically show how to implement a stroboscopically repeated, near-projective measurement of a photon impinging on one of the cavities. Optimizing rate, duration, and strength of the measurement by flux control of the junction 
and developing a threshold protocol to detect the photon from a homodyne measurement of the radiation output of the other cavity,  we achieve highly efficient detection with low dark counts. 
By cascading the detector with a preamplifier, where a similar two-cavity Josephson-photonics device acts as a photon multiplier, we can further improve the device to reach a detection efficiency of $88.5 \%$ with a dark count rate of 
$\sim10^{-4} \gamma_a$, set by the resonance width $\gamma_a$ of the absorbing cavity. These results for a multiplication factor of two suggest that near-unity efficiencies may be reached for higher multiplication factors. 
\end{abstract}
\maketitle

\section{Introduction}
The efficient detection of single itinerant microwave photons remains a highly challenging task. This functionality is a key ingredient for many protocols and technologies in quantum information processing, quantum sensing, and quantum communication that 
rely on the quantum dynamics of superconducting electronic devices \cite{casariego_propagating_2023, gu_microwave_2017}.
More recently, they have also been used in magnetic resonance spectroscopy on single spins \cite{wang_single-electron_2023,travesedo_all-microwave_2025,osullivan_individual_2025} or spin ensembles \cite{albertinale_detecting_2021, billaud_electron_2025} and in dark matter search experiments 
\cite{dixit_searching_2021, braggio_quantum-enhanced_2025, pankratov_towards_2022}.
The requirements for different applications are as varied as the schemes of proposed and realized single-photon microwave detectors (see recent reviews \cite{casariego_propagating_2023,gu_microwave_2017,sathyamoorthy_detecting_2016} and references within).
Schemes include bolometric detectors \cite{karimi_quantum_2020,karimi_bolometric_2024,kokkoniemi_nanobolometer_2019,kokkoniemi_bolometer_2020,gunyho_single-shot_2024}, which have not yet reached single-photon sensitivity, threshold detectors \cite{chen_microwave_2011,ladeynov_detection_2025,pankratov_towards_2022,rettaroli_josephson_2021,oelsner_detection_2017,he_experimental_2025, wang_observing_2025}, efforts to improve efficiency by arrays or cascades of detectors \cite{fan_nonabsorbing_2014, sathyamoorthy_quantum_2014, 
royer_itinerant_2018,navez_quantum_2023}
, various platforms incorporating qubits driven by complex pulse sequences \cite{inomata_single_2016,besse_single-shot_2018,kono_quantum_2018}, or devices specifically designed for resolving the photon-number distribution of a pulse with known arrival time \cite{dassonneville_number-resolved_2020}. 

Many proposals incorporate as the first step in the detection chain a cavity, on which the itinerant photon impinges. The state of the cavity is measured to detect if a photon has been absorbed, thus rendering the task of detecting an itinerant photon into the read-out of a quantum state. It has long been recognized
\cite{helmer_quantum_2009, royer_itinerant_2018}
that in any such proposal the measurement of the cavity's state induces a Zeno-type backaction, which adversely affects the absorption of the incoming photon by the cavity, limiting the detection efficiency.

Here, we revisit this basic idea, but introduce three novel ingredients to such a scheme:
Firstly, we show within a toy-model, that  stroboscopically repeated, instantaneous projective measurements of the cavity, which leave the cavity free to absorb the photon undisturbed in between the measurement instances, can achieve efficient detection.
Secondly, we show that such a scheme can be realized utilizing existing \emph{Josephson-photonics devices}, where two microwave resonators are coupled by a dc-voltage biased Josephson junction. These versatile devices can implement a variety of unconventional nonlinear single- or multi-photon drives \cite{gramich_coulomb-blockade_2013, armour_universal_2013, leppakangas_nonclassical_2013, leppakangas_antibunched_2015, 
armour_josephson_2015, trif_photon_2015, aissaoui_cat_2024} to create microwave radiation with interesting (quantum) properties \cite{hofheinz_bright_2011, westig_emission_2017, chen_realization_2014, cassidy_demonstration_2017, rolland_antibunched_2019, grimm_bright_2019, peugeot_generating_2021, menard_emission_2022}, and have also been proposed \cite{leppakangas_multiplying_2018, danner_amplification_2025} and tested \cite{jebari_near-quantum-limited_2018, albert_microwave_2024, martel_influence_2025} for amplification and detection tasks. Here, they are used to implement a drive of the second cavity which (i) depends on the state of the first one and (ii) can be switched on and off via the flux bias of a superconducting quantum interference (SQUID) Josephson junction. Sending the itinerant photon on the first and monitoring the output of the second cavity by homodyne detection then mimics the aforementioned toy-model. Investigating the full simulation of the device's quantum dynamics under a detailed measurement protocol, we find promising results with detection efficiency reaching $69.8 \%$ for realistic system parameters.
Thirdly and finally, 
 we will describe how our scheme can be improved by preamplifying the photon with the same type of Josephson-device, acting as a photomultiplier as described in \cite{leppakangas_multiplying_2018, danner_amplification_2025, albert_microwave_2024}. 
Simulating a cascaded chain of a two-fold multiplier and a detector, we will show a detection efficiency which closely approaches simple estimates based on the toy-model, promising further improvements for higher multiplication factors.

\section{Stroboscopic projective detection of an itinerant photon} \label{sec:Proj}

\begin{figure}[t]
    \centering
\includegraphics[width=\columnwidth]{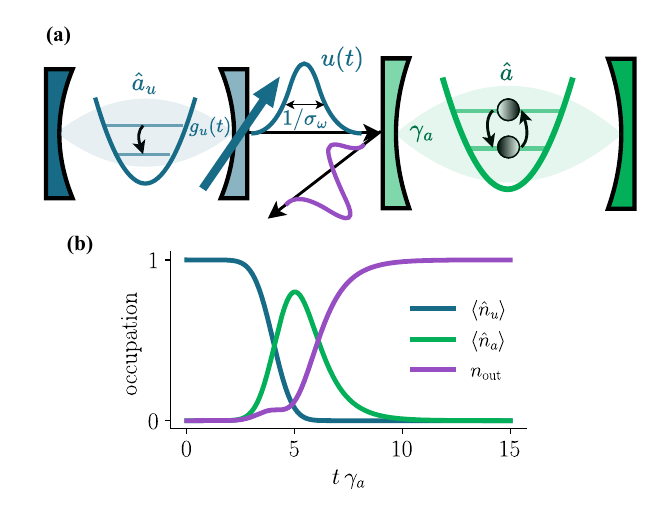}
    \caption{
    (a) Effective description of a microwave cavity driven by an input field. The scenario is theoretically modeled by a cascaded Master equation involving an auxiliary cavity $a_u$ that emits the quantum pulse $u(t)$ 
    via a tunable decay coupling $g_u(t)$ (blue arrow). We assume a Gaussian mode 
    $u(t) \sim \ee^{-(\sigma_\omega t)^2/2}$
    with spectral width $\sigma_\omega=\gamma_a$ in a single-photon Fock state $\ket{\psi}_u=\ket{1}_u$. The input field is partly reflected and partly absorbed and re-emitted by the microwave cavity. 
    (b) Occupations of auxiliary cavity $a_u$ emitting the photon, cavity $a$ which absorbs and subsequently emits the photon, and the integrated output $n_\mathrm{out}=\int_0^t \avg{\loss_a^\dagger \loss_a}(t') \dd t'$. 
    The plateau in $n_\mathrm{out}$ at $t\gamma_a \approx 4$ indicates destructive interference of contributions from direct reflection and from photon absorption and re-emission.}
    \label{fig:input_and_cav}
\end{figure}

\begin{figure}[b]
    \centering
    \includegraphics[width=\columnwidth]{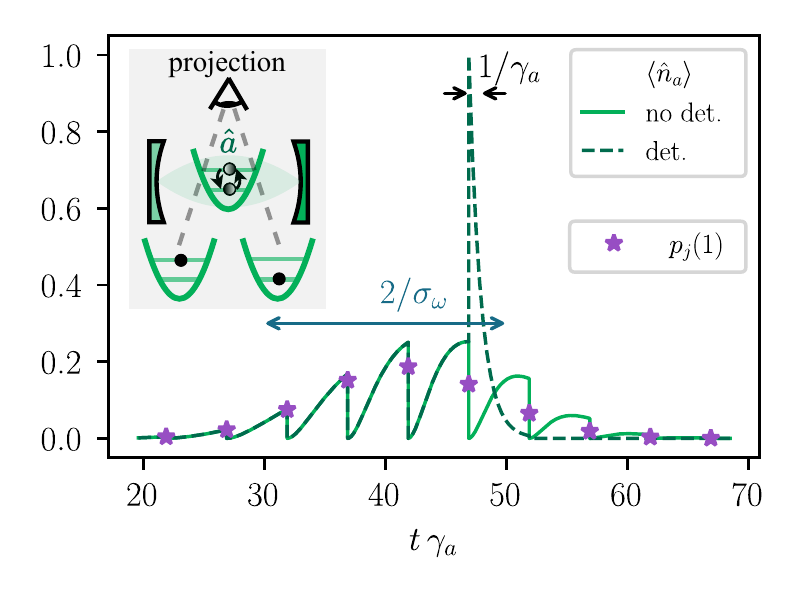}
    \caption{
    Stroboscopic projective measurements (performed with frequency $\gamma_\mathrm{m}=0.2 \gamma_a$) of a microwave cavity $a$ that is subject to an impinging resonant Gaussian single-photon pulse ($\sigma_\omega/\gamma_a=0.1$). Each measurement projects the cavity's state either to $\ket{1}_a$ (i.e.~the photon is detected) or to $\ket{0}_a$, where the photon remains undetected (c.f.~schematic setup in the inset). The solid line shows a trajectory where the photon was never detected, while the dashed trajectory detects the photon in the measurement at $t_{\mathrm{m},j} =47/\gamma_a$, where the (total) state is projected to $\ket{\psi}=\ket{01}_{u,a}$. Because of the subsequent exponential decay, the cavity occupation will be (with very high probability) projected to vacuum in the following measurement. The probability to find the photon at measurement $j$, $p_j(1)=p_j(O_j=1,\,O_k=0 \,\forall\, k<j)$ is indicated by purple markers, while the total photon detection probability is $\eta=67\%$. The y-axis shows both the occupation and the probability (same scale). 
    [Parameters: $\gamma_\mathrm{m}=0.2\gamma_a$, $\sigma_\omega=0.1\gamma_a$, $\Delta_a=0$.]
    }
    \label{fig:project_two_traj}
\end{figure}

\begin{figure}[b]
    \centering
    \includegraphics[width=\columnwidth]{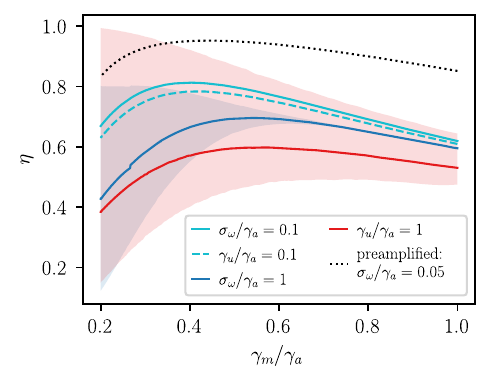}
    \caption{
    Detection probability $\eta$ depending on the measurement rate $\gamma_\mathrm{m}$ for different pulse shapes. 
    The curves were averaged over several measurement positions relative to the pulse. Shaded regions indicate the range of detection probabilities depending on the relative measurement position. \\
    A maximum detection probability of $\eta=81 \%$ can be achieved for a measurement rate $\gamma_m=0.4\gamma_a$ for spectrally narrow and resonant Gaussian input pulses (solid light blue, $\sigma_\omega=\gamma_a/10$), while it peaks at  $\eta=78 \%$ 
    for resonant exponential input pulses, $u(t)= \sqrt{\gamma_u} \, \theta(t) \,  \ee^{-\gamma_ut/2}$ (where $\gamma_u=\gamma_a/10$, dashed). \\
    For spectrally broader Gaussian input pulses ($\sigma_\omega=\gamma_a$, dark blue line and region), and for a time-reversed exponential pulse $u(t)= \sqrt{\gamma_u} \, \theta(-t) \,  \ee^{\gamma_ut/2}$ ($\gamma_u=\gamma_a$, red line and region), the maximum detection probability is significantly smaller. Due to the stricter time confinement of those pulses, the position of the measurements relative to the pulses' arrival time strongly influences the total detection probability, especially for low measurement frequencies.  
    A preamplification of a resonant single-photon Gaussian input pulse ($\sigma_\omega=\gamma_a/20$) with multiplication factor $n=2$ (c.f.~Sec.~\ref{sec:MultiCasc}) enables near-deterministic detection, $\eta=95 \%$.}
    \label{fig:project_opt}
\end{figure}

As explained in the Introduction this paper will present a novel device enabling the detection of a single itinerant microwave photon by implementing a stroboscopically repeated, near-projective measurement of the occupation of a cavity, on which the itinerant photon impinges. 
If the photon is absorbed (as opposed to reflected) by the cavity, 
the projective measurement gathers information on the photon's existence. 
The described measurement process is non-destructive in the sense that subsequent to the measurement the photon decays from the cavity to its output transmission line and could, in principle, be subjected to further measurements~\footnote{Note, however, that the strong scrambling of the photonic mode structure makes a simple cascading impossible, as we will discuss below.}.
Before presenting the experimental platform and the complete protocol in full, it is instructive to briefly investigate a simplified toy model that assumes a device-independent perfect projective measurement. 

To start with, we need a full quantum description of the deterministic dissipative dynamics of a single photon impinging on a cavity and of the probabilistic quantum measurement process. The latter will then be replaced by the more complex description of the actual implementation of the measurement device in Secs.~\ref{sec:JPD-Proj} and \ref{sec:MultiCasc}. 

A single bosonic input field probing a quantum system is characterized by its carrier frequency $\omega_u$, its incident time-dependent pulse shape $u(t)$, and by its quantum state  (either a pure state $\ket{\psi}_u$ or a mixed state $\hat{\rho}_u$).
We specifically consider the experimental scenario of a few-photon pulse impinging on a single microwave cavity with (rotating-frame) Hamiltonian $\hat{H}_0=\hbar\Delta_a \hat{n}_a$ with loss rate $\gamma_a$, where $\Delta_a=\omega_a-\omega_u$ is the rotating-frame detuning.
Theoretically, that situation is described following the framework presented by the M{\o}lmer group~\cite{kiilerich_input-output_2019,kiilerich_quantum_2020}. The approach relies on eliminating the explicit description of the propagating field by the introduction of an auxiliary cavity $a_u$ with a time-dependent loss-rate $g_u(t)$, see Fig.~\ref{fig:input_and_cav}(a), specifically constructed to emit the quantum pulse $u(t)$ \cite{kiilerich_input-output_2019,kiilerich_quantum_2020}. A cascaded Master equation with Hamiltonian $\hat{H}=\hat{H}_0 + \left( \frac{\ii \hbar}{2} g_u(t)\sqrt{\gamma_a} \,  \adag_u \ann +\mathrm{h.c.} \right)$ and single Lindblad operator $\loss_a=g_u^*(t) \ann_u + \sqrt{\gamma_a} \ann$ ensures the uni-directional influence of the quantum pulse driving the cavity.
To demonstrate this effective description, the occupations of $a_u$ and $a$ as well as the integrated output of the system, when a photon in a resonant Gaussian mode impinges on $a$ are shown in Fig.~\ref{fig:input_and_cav}(b). 
The photon is emitted from $a_u$ (blue), impinges on cavity $a$ where it is partly reflected and partly absorbed and re-emitted (green) to the environment, where the integrated output finally reaches unity (violet).

Projective measurements performed on the cavity $a$ are added by simulating the system up to the point of a projective measurement at $t_{\mathrm{m},j}$ with the same Lindblad master equation for $a_u$ and $a$. Then, depending on the outcome $O_j \in \{0,1\}$ of the measurement $j$, the new state $\dens'$ of the system is given by 
\begin{equation}
    \dens'(t_{\mathrm{m},j}|O_j)=\frac{\proj_{O_j} \dens(t_{\mathrm{m},j}) \proj_{O_j}}{\mathrm{tr}\left[ \proj_{O_j}\dens(t_{\mathrm{m},j}) \proj_{O_j} \right]}
\end{equation}
with $\proj_{1}=\mathds{1}_u \otimes \ketbra{1}{1}_{a}$ when a photon is detected, and $\proj_{0}=\mathds{1}-\proj_{1}$ if not. Subsequently, the state $\dens'$ is evolved with the Lindblad master equation up to the next measurement. 

Fig.~\ref{fig:project_two_traj} shows two distinct trajectories with stroboscopically repeated projective measurements. 
When a measurement finds no photon as for all measurements of the first trajectory (solid line), the cavity occupation, $\occ_a$, resets to $0$ and has to be driven up by the incoming photon pulse again. 
When a photon is found as in the measurement at $t=47/\gamma_a$ of the second trajectory (dashed line), the state of the system is projected to $\ket{01}_{u,a}$. Afterwards, the photon decays from the cavity $a$ to the environment. As the quantum pulse does not contain any photons after the detection of the photon in $a$, the pulse does not drive cavity $a$ thereafter.  

The probability to miss the photon in measurement $j$, conditional on the previous measurement outcomes, is $p_j(O_j=0|\{O_k\}, k<j) = \mathrm{tr}\left[ \dens(t_{\mathrm{m},j}) \proj_0 \right]$. Thus, the photon detection probability $\eta$ of finding the photon in any of the $M$ measurements, is given by $\eta=1- \Pi_{j=0}^M \, p_j(0|O_k=0 \, \forall \,  k<j) \, $.
That overall detection probability $\eta$ depends on the pulse shape $u(t)$, the measurement rate $\gamma_\mathrm{m}=1/t_\mathrm{dist}$, where $t_\mathrm{dist}$ is the time between measurements, and the  position of the measurements relative to the quantum pulse. The dependence of $\eta$ on $\gamma_\mathrm{m}$ is depicted in Fig.~\ref{fig:project_opt} for several pulse shapes (Gaussian, rising/decaying exponential). 
To detect a photon with high probability, it has to be resonant in order to enter the cavity efficiently (cf. the solid curves in light/dark blue for Gaussian pulses with small/large frequency spread), and measurements should occur frequently enough, so that it is not missed. 
However, measuring too frequently freezes the cavity in the empty state by the quantum Zeno effect: the initially quadratic increase in occupation after an unsuccessful measurement
leads to a very small probability for a measurement following shortly after. 
This decrease in probability cannot be compensated by the 
increased number of measurements. In accordance with these simple arguments, we find an optimum at a moderate detection frequency of $\gamma_\mathrm{m} \approx 0.4 \gamma_a$  (for resonant pulses). 

Given resonant pulses, their exact pulse shape is less important, as exemplified by comparing the solid light blue to the dashed light blue line in Fig.~\ref{fig:project_opt}; the latter stemming from the (natural) exponential decay of an input cavity with constant decay rate. 

It is also instructive to consider a single-photon pulse shaped to the time-reversed version of the natural exponential decay, that has recently raised attention, as it can be perfectly absorbed \cite{wenner_catching_2014}. A cavity with matching decay rate will reach unity occupation at the time the pulse stops, see App.~\ref{app:rev_exp_pulse}, before it eventually decays. Were a single measurement to occur at that time of unity occupation, it would result in deterministic detection (upper left corner of the red shaded region in Fig.~\ref{fig:project_opt}). However, if the arrival time of the photon is unknown, the resulting detection probabilities have to be averaged over all possible positions of the stroboscopic measurements (shaded vertical range), and the resulting averaged probability (solid red line) is quite small for that pulse shape. 

\section{A Josephson-photonics device implementing projective detection} \label{sec:JPD-Proj}

\begin{figure}[b]
    \centering
    \includegraphics[width=\columnwidth]{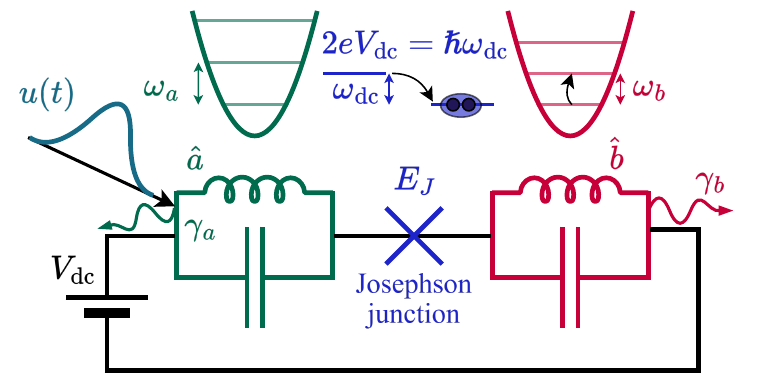}
    \caption{
    Schematic electrical circuit of a Josepshon-photonics device consisting of two microwave cavities connected in series with a dc-biased Josephson junction. By tuning the voltage to the single-photon resonance of cavity $b$, $\omega_\mathrm{dc}=\omega_b$, each Cooper pair (CP) tunneling through the junction effectively creates one photon in $b$. The linear driving of cavity $b$ is renormalized by virtual excitations [c.f.~the Bessel functions in Eq.~\eqref{eq:RWA_H}], and thereby depends on the state of cavity $a$. An impinging quantum pulse modifies the state of cavity $a$, which effectively modulates the driving and thus the quantum state of cavity $b$. 
    }
    \label{fig:JPD}
\end{figure}

\begin{figure}[b]
    \centering
    \includegraphics[width=\columnwidth]{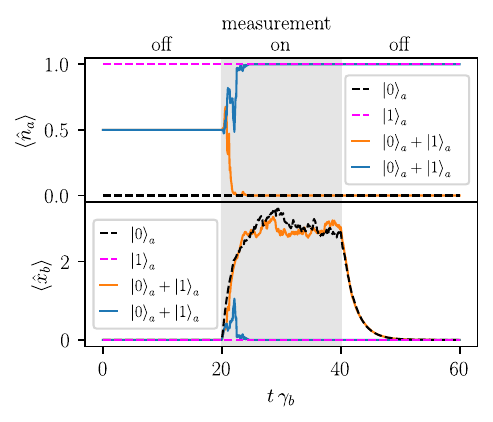}
    \caption{
    Measuring a closed cavity $a$, $\gamma_a=0$, in a Josephson-photonics setup.
    The coupling $E_J$ is switched on between $t\gamma_b=20$ and $t \gamma_b=40$. The top panel shows the occupation of cavity $a$, while the bottom panel shows the quadrature $\avg{\hat{x}_b}$ of cavity $b$. 
    If cavity $a$ is empty, cavity $b$ is driven to a (near-)coherent state $\ket{\beta=1.6}$ and returns to vacuum when $E_J$ is switched off (dashed black lines). 
    For $\alpha_0=1$ and when $\ket{\psi}_{0,a}=\ket{1}$, the effective driving strength of $b$ vanishes, such that $b$ stays in vacuum irrespective of the coupling (dashed magenta lines).
    When cavity $a$ starts in a coherent superposition $\ket{\psi}_{0,a}=(\ket{0}+\ket{1})/\sqrt{2}$, the measurement projects its state to either $\ket{0}_a$ (solid orange lines) or $\ket{1}_a$ (solid blue lines) with a probability of $1/2$ each. \newline
    [Parameters: $\gamma_b=1$,  $\beta_0=0.3$, $\Delta_a=\Delta_b=0$, $E_J^* \beta_0/(\hbar \gamma_b)=1.6$.]
    }
    \label{fig:closed_a}
\end{figure}

\begin{figure}[b]
    \centering
    \includegraphics[width=\columnwidth]{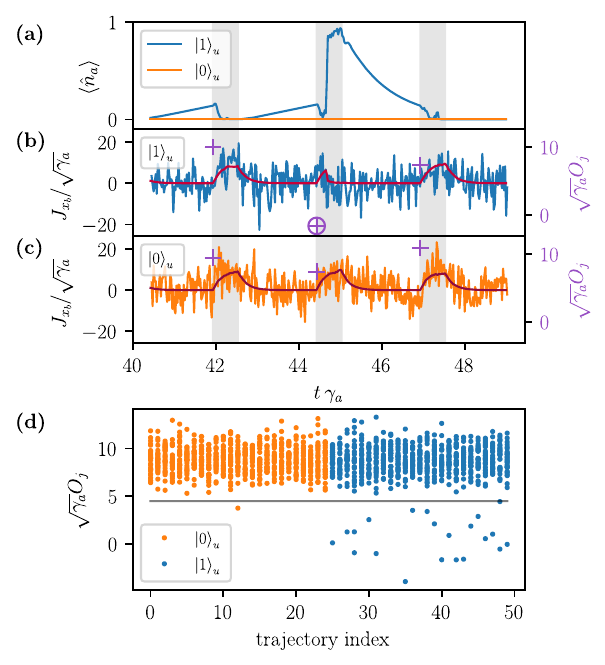}
    \caption{
    (a) Occupation $\occ_a$ with (blue) and without (orange) resonant Gaussian photon input for three consecutive measurements (grey areas). Absorption of the photon happens in the second measurement shown, where $a$ is projected to $\ket{1}$. 
    (b)/(c) Measurement outcome $J_{x_b}$ (blue/orange) and expectation value $\sqrt{\gamma_b} \langle\hat{x}_b\rangle$ without noise (red) corresponding to the two trajectories 
    shown in (a). 
    The outcomes $O_j$ of the three measurements are plotted as purple crosses (scale on the right). 
    The measurement where a photon is detected shows a significantly smaller outcome $O_j$ (circled cross).
    (d) Measurement outcomes for several trajectories with/without (blue/orange) input photon, where each trajectory was measured $31$ times (vertical stacks). 
    Data points below a chosen threshold (gray line) are classified as detection events.\newline
    [Parameters: $\gamma_\mathrm{m}/\gamma_a=0.4$, $t_\mathrm{m} \gamma_a=0.6$, $\sigma_\omega/\gamma_a=0.1$, $t_0 \gamma_a =40$, $\gamma_b/\gamma_a=10$, $\alpha_0=1$, $\beta_0=0.3$, $\Delta_a=\Delta_b=0$, $E_J^* \beta_0/(\hbar \gamma_b)=1.6$.]
    }
    \label{fig:threshold}
\end{figure}

Aiming to implement the projective measurement of the occupation of a cavity $a$, discussed in its idealized abstract version in the previous section, we use the two-cavity Josephson-photonics device (JPD) shown in Fig.~\ref{fig:JPD}.  
Two cavities are connected in series with a dc-voltage biased Josephson junction, such that the system can be described by the Hamiltonian~\cite{armour_josephson_2015} 
\begin{equation}\label{eq:full_H}
    \begin{split}
    \ham_\mathrm{JPD}=&\hbar \omega_a \adag \ann + \hbar \omega_b \bdag \bnn \\
		&- E_J \cos[\omega_\mathrm{dc} t +\alpha_0 (\adag+\ann) +\beta_0 (\bdag+\bnn ) ]  \, ,
        \end{split}
\end{equation}
where the dc-voltage can be associated with a driving frequency $\omega_\mathrm{dc}=2eV_\mathrm{dc}/\hbar$.
In such a device, the tunneling of a Cooper pair across the junction (represented by the $\ee^{\pm i \omega_\textrm{dc}t}$ terms in the Hamiltonian) couples to processes containing any number of photon creation/destruction events in the two cavities, $ \left[\alpha_0\hat{a}^\dagger\right]^n \left[\alpha_0\hat{a}\right]^m \; \left[\beta_0\hat{b}^\dagger\right]^s\left[\beta_0\hat{b}\right]^t\;(m,n,s,t \in \mathbb{N}_0) $, stemming from the expansion of the $\cos$-term in Eq.~\eqref{eq:full_H}, where $\alpha_0, \beta_0$ are the zero-point fluctuations of the phase variables of cavities $a,\,b$ \cite{gramich_coulomb-blockade_2013, armour_universal_2013, armour_josephson_2015}. 
Relevant are those terms, where the energy of the tunneling Cooper pair closely matches the energy of created and destroyed photons. For instance, a pair of entangled photons in cavity $a$ and $b$ may be created, if $\omega_\textrm{dc,ent} = \omega_a + \omega_b$~\cite{westig_emission_2017, dambach_generating_2017,peugeot_generating_2021}. 
Here, we tune the dc-bias to $2 e V_\mathrm{dc} = \hbar \omega_\mathrm{dc} \approx \hbar\omega_b$ such that there is a dominant low-order term, $\cos{( \omega_\textrm{dc}t)} \beta_0 (\hat{b}^\dagger+\hat{b})$, driving cavity $b$ linearly, while cavity $a$ appears as a mere bystander to this process. 
However, from the argument above it is obvious that this simple process is modified by other resonant processes involving arbitrary powers of $\hat{n}_a=\hat{a}^\dagger \hat{a}$ and $\hat{n}_b$ which (i) make the single-photon driving of $b$ nonlinear, and, more importantly for this work, (ii) make the drive of cavity $b$ depend on the occupation of cavity $a$.

To isolate the so-renormalized resonant process, we move to the rotating frame and apply a rotating-wave approximation (RWA) yielding 
\begin{equation}
\label{eq:RWA_H}
    \begin{split}
        \ham_\mathrm{RWA}=&\hbar \Delta_a \adag \ann + \hbar \Delta_b \bdag \bnn \\
        &- \ii  \frac{E_J^* \beta_0}{2} \normord{\left[  \bnn- \bdag \right]
		\mathrm{J}_0\left(2\alpha_0 \sqrt{\hat{n}_a} \right) \frac{\mathrm{J}_1\left(2\beta_0 \sqrt{\hat{n}_b} \right)}{ \beta_0 \sqrt{\hat{n}_b}}  } \\
        \approx& \hbar \Delta_a \adag \ann + \hbar \Delta_b \bdag \bnn - \ii  \frac{E_J^* \beta_0}{2} \left[  \bnn- \bdag \right]
		\left(1-\alpha_0^2 \occ_a \right) 
        \end{split}
\end{equation}
with $\Delta_a=\omega_a-\omega_{a,\mathrm{rot}}$ (usually set by the input pulse, i.\,e.\,, $\omega_{a,\mathrm{rot}}=\omega_u$), $\Delta_b=\omega_b-\omega_\mathrm{dc}$ and $E_J^*=E_J \ee^{-(\alpha_0^2+\beta_0^2)/2}$. 
The approximation leading to the last line assumes negligible nonlinearities in the driving of $b$ (i.e.~$\beta_0\sqrt{\langle\hat{n}_b\rangle}\ll1$) and a single photon in cavity $a$, for which case the Bessel function, $\mathrm{J}_0\left(2\alpha_0 \sqrt{\hat{n}_a} \right)$, can be projected to the $n_a=0,1$ subspace.
The moment-occupation coupling $\hat{p}_b \occ_a$ realizes a driving strength for cavity $b$ which depends on the state of cavity $a$.
As direct consequence, a photon detection scheme can be devised that monitors observables of $b$, such as homodyne or heterodyne quadrature measurements, and thereby acquires information on the occupation of $a$.
In the following, we will lay out how this basic idea is extended to a full scheme of stroboscopic detection of itinerant photons. First, we will sketch how a single near-projective measurement of the state of a closed ($\gamma_a \equiv 0$) cavity $a$ is realized, by switching on a strong Josephson coupling, $E_J$, and thus an $\hat{n}_a$-dependent driving of cavity $b$ in the JPD setup of Fig.~\ref{fig:JPD} \footnote{The effective Josephson coupling can be tuned by a magnetic flux, if the junction is realized as a superconducting quantum interference device (SQUID).}. 
To infer the information on $\hat{n}_a$, a convenient measurement observable has to be extracted from the monitored output of $b$. After characterizing a single measurement, we will turn in Sec.~\ref{sec:JPDStrobo} to stroboscopically-repeated fast measurements of an open cavity $a$, on which a single-photon pulse impinges.

\subsection{Homodyne detection}
\label{sec:homodyne_detection}

We want to implement a near-projective measurement of cavity $a$ by monitoring the $x$-quadrature of the microwave-radiation output of cavity $b$. 
To describe such a homodyne measurement, we follow the standard quantum-optics approach \cite{wiseman_quantum_2009} and employ the stochastic master equation, 
\begin{equation}
	\dd \dens(t)= d_1 \dens \, \dd t +  d_{2} \dens \, \dd W \ , 
\end{equation}
with the Wiener increment 
\begin{equation}
	\avg{\dd W}= 0 \ , \quad \avg{\dd W^2}=\dd t\ . 
\end{equation} 
The \emph{deterministic} part of the master equation is given by 
\begin{equation}
	d_1 \dens =-\ii [\ham,\dens] +  \mathcal{D}[\hat{C}] \dens +  \mathcal{D}[\hat{S}] \dens \ , 
\end{equation}
with 
\begin{equation}
	\mathcal{D}[\loss] \dens =\frac{1}{2} \left[ 2 \loss \dens \loss^\dagger - \dens \loss^\dagger \loss - \loss^\dagger \loss \dens \right] \ . 
\end{equation}
For our set-up, the unmonitored operator is $\hat{C} \equiv \loss_a = \sqrt{\gamma_a} \ann \;\; (\;+\, g_u^*(t) \ann_u\;)$, see above, 
as we do not observe cavity $a$ directly, while the output of $b$ is monitored, $\hat{S} \equiv \loss_b=\sqrt{\gamma_b} \bnn$. 
The Hamiltonian $\ham$ is the full model Hamiltonian, potentially including the cascaded auxiliary cavity. 

The \emph{stochastic} part is calculated by 
\begin{equation}
	d_{2} \dens = \hat{S} \dens + \dens \hat{S}^\dagger - \mathrm{tr}\left[ \hat{S} \dens + \dens \hat{S}^\dagger\right] \dens \ . 
\end{equation}
The stochastic quantum master equation then yields individual trajectories for the quadrature measurement, 
\begin{equation}\label{eq:noisy_quadrature}
	J_{x_b}=\avg{\loss_b+ \loss_b^\dagger} + \dv{W}{t} \ , 
\end{equation}
where the derivative of the Wiener increment simulates vacuum noise associated with each  measurement run. These trajectories have the same  noise properties as an experiment, where the radiation leaking out of $b$ is fed into a phase-sensitive quantum-limited amplifier, whose output is subsequently recorded and processed classically. 

To implement a projective measurement of cavity $a$, the Josephson coupling $E_J$ is switched on, so that cavity $b$ is driven on its fundamental single-photon resonance, $\Delta_b=\omega_b-\omega_\textrm{dc}=0$, assuming a perfectly stable dc-voltage for the purpose of this publication \footnote{
In fact, it is known that voltage fluctuations lead to a slow diffusion of the effective phase of the driving (and the corresponding cavity state in phase space) and how this can be overcome by phase locking \cite{danner_injection_2021, hohe_quantum_2025, danner_quantum_2025}. For experimental realizations of the detection scheme, we rather envision that we can either monitor a slow phase diffusion and correct the phase of the homodyne quadrature measurement, or straightforwardly adapt our scheme to heterodyne measurement of both quadratures.}. 
According to Eq.~\eqref{eq:RWA_H}, if cavity $a$ is in vacuum, cavity $b$ is driven into the coherent state $\ket{\beta}$ with $\beta=E_J^* \beta_0/( \hbar\gamma_b)$. The effective driving of $b$ is reduced when cavity $a$ is occupied by a photon and, indeed, vanishes for  $\alpha_0=1$. The results of such measurements on a closed (i.e.~$\gamma_a=0$) cavity $a$ prepared in different initial states are shown in Fig.~\ref{fig:closed_a}.  After $E_J$ is switched on, the quadrature $\langle \hat{x}_b\rangle$ (shown in Fig.~\ref{fig:closed_a} for individual trajectories without the Wiener increment) quickly approaches a finite value or zero, while the state of the cavity is projected to the corresponding measurement-eigenstates $\ket{0}$ or $\ket{1}$, if the initial state is a superposition $\ket{\psi}_a=(\ket{0}+\ket{1})/\sqrt{2}$.

\subsection{Stroboscopically-repeated measurements }\label{sec:JPDStrobo}
Keeping the Josephson coupling constantly on, $E_J\neq0$, is not conducive to our ultimate goal of detecting itinerant single-photon pulses, as it will strongly reduce the probability that the incoming photon is absorbed by the cavity. This can be seen as being due to a strong dephasing noise acting on cavity $a$ caused by the fluctuations of the state of cavity $b$ [c.f.~the
$\left(  \bnn- \bdag \right) \occ_a$ term in the Hamiltonian \eqref{eq:RWA_H}] or as an effect analogous to the quantum-Zeno-like reduction of the detection probability for the regime of projective measurements with high rate in Fig.~\ref{fig:project_opt}. 
In consequence, we turn to stroboscopically switching $E_J$ on only for a short measurement interval, just long enough to determine the state in $a$, and repeat such near-projective measurements with a frequency similar to the optimal frequency for our toy-model's projective measurements in Fig.~\ref{fig:project_opt}. This requires a separation of time scales, $\gamma_b\gg \gamma_a$, so that $b$ adapts to the state in $a$ very quickly and thus mimics the ideal instantaneous projective measurement. 

Fig.~\ref{fig:threshold}(a--c) gives two example trajectories, one with and one without an incoming photon, where three consecutive measurements are performed in the manner just described. Shown are the occupation of cavity $a$ [in Fig.~\ref{fig:threshold}(a)] and in (b--c) the measurement signal, the observed quadrature $\langle \hat{x}_b \rangle$ with the explicit noise term of Eq.~\eqref{eq:noisy_quadrature} (blue and orange), and for comparison without that term (solid red lines).

Considering first the trajectory without photon input (dark green line in (a) and (c)), we see how cavity $b$ is driven towards its finite-quadrature steady state during the measurement, while relaxing back to the ground state when $E_J$ is off again. With a single-photon input, similar behavior of cavity $b$ is observed for the first and third measurement shown, where the state in $a$ is projected to $ \approx \ket{0}$ as seen in (a). For the second measurement on the other hand, cavity $a$ is projected to $ \approx \ket{1}$ and the quadrature of $\langle \hat{x}_b \rangle$ remains closer to $0$. However, the noise on the actual observed $J_{x_b}$, makes the distinction between measurement outcomes harder. 

To distinguish the difference of the signals hidden in background noise, $J_{x_b}$ is integrated 
\begin{equation}\label{eq:Oi}
    O_j=\int_{t_{j, \mathrm{start}}}^{t_{j+1, \mathrm{start}}} J_{x_b}(t) w(t) \dd t
\end{equation}
with a weight function $w(t)$ to obtain a single real value $O_j$ for each measurement [indicated by $+$-markers (b) and (c) of Fig.~\ref{fig:threshold}]. Low values of $O_j$ as in the second measurement of the middle panel indicate a successful photon detection.
The \emph{weight function} $w(t)$, shown and discussed in more detail in Appendix~\ref{app:weight}, disregards values at the start and long after the measurement, where the expected observed quadrature of $b$ does not depend on the state of $a$, while peaking where that dependence is maximal.\\
For the \emph{classification of measurements} a  threshold is defined, below which a photon detection event is declared. 
Measurement outcomes $O_j$ for some trajectories  without (orange points on the left) and with input (right) are shown in Fig.~\ref{fig:threshold}(b) with a threshold line set to classify as many measurements correctly as possible. 

With the chosen circuit parameters, weight function, and threshold definition we obtain a detection efficiency of $\eta=69.8 \%$ coming quite close to detection probability of the idealized projective scheme, which reached $81 \%$. The probability to wrongly identify a photon in one measurement is $0.029 \%$, which results in a dark rate of $\gamma_\mathrm{dark}=0.116 \cdot 10^{-3} \gamma_a$ assuming a mean rate of $ 12.5 \cdot 10^{-3} \gamma_a$ incoming photons. Dark counts can be curtailed at the expense of reduced detection efficiency by lowering the threshold value (see also Fig.~\ref{fig:epsilonGammaDark} below).
The comparison with the projective measurement toy-model suggest, that improved efficiency at low dark counts may be approached by faster measurements, $\gamma_b/\gamma_a \rightarrow \infty$ and a clearer distinction of measurement outcomes, i.e., a larger displacement $\beta$ between the respective steady states of cavity $b$ for empty and occupied cavity $a$. These conditions imply an increase of the Josephson coupling $E_J$ which might lead to the breakdown of the RWA approximation and the validity of the corresponding Hamiltonian Eq.~\eqref{eq:RWA_H}. 
Effects emerging from rotating terms of the Hamiltonian will cause spontaneous excitations in $a$, drastically increasing the dark rate. Choosing the resonance frequencies of the cavities and other parameters carefully, we can estimate, see Appendix~\ref{app:RWA}, that this additional dark rate stays at a similar magnitude as the one found within RWA, so that the general functionality of the device remains at the cost of a reduced operating bandwidth. Furthermore, it is known how far-off resonant non-RWA effects can be curbed by designing the impedance and shape the oscillator resonances  \cite{nehra_dc-powered_2025}. 

\section{Preamplification for improved detection} \label{sec:MultiCasc}

\begin{figure*}[t]
    \centering
    \includegraphics[width=\textwidth]{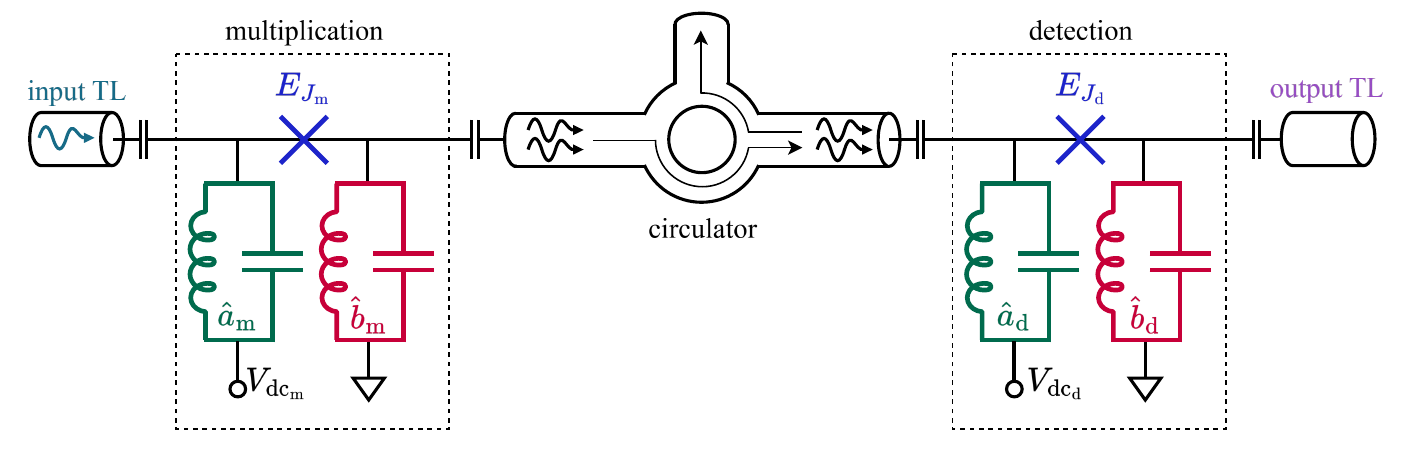}
    \caption{Schematic detection setup with preamplification by a Josephson-photonics device that works as an inelastic Cooper pair tunneling photon multiplier \cite{danner_amplification_2025}. 
    One photon impinging from the input transmission line (TL) is absorbed by cavity  $a_\mathrm{m}$ and multiplied to $n$ photons in cavity $b_\mathrm{m}$ by a tunneling Cooper pair when the dc bias fulfills $\omega_{\mathrm{dc}_\mathrm{m}}+\omega_{a_\mathrm{m}} \approx n \omega_{b_\mathrm{m}}$ (here $n=2$). 
    These photons are sent as input into the detection device as described in Sec.~\ref{sec:JPD-Proj}. 
    The circulator ensures that there is no back-scattering from the measurement device on the multiplier. 
    }
    \label{fig:MultiCascSpec}
\end{figure*}

\begin{figure}[b]
    \centering
    \includegraphics[width=\columnwidth]{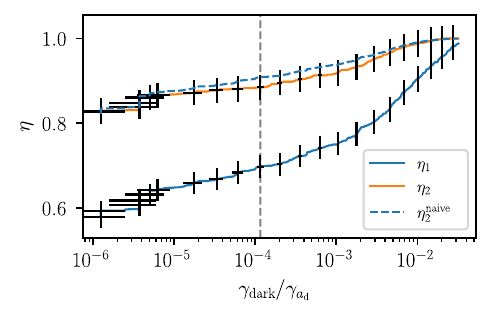}
    \caption{Detection efficiency $\eta$ depending on the dark rate $\gamma_\mathrm{dark}$ for the simple scheme from Sec.~\ref{sec:JPDStrobo} (blue) and with the preamplificiation discussed in Sec.~\ref{sec:MultiCasc} (orange). The dashed blue line shows the naive expectation for two independent photons.
    From left to right, the threshold is increasing, $\sqrt{\gamma_{a_\mathrm{d}}} O_\mathrm{th}=2.7\hdots7$. For parameters see Fig.~\ref{fig:threshold} for the detection JPD and App.~\ref{app:multiplier} for the preamplifier. 
    The dashed grey line indicates the threshold chosen in Fig.~\ref{fig:threshold}(d) to obtain $\eta$ and $\gamma_\mathrm{dark}$ in the main text. 
    The Gaussian input pulse has width $\sigma_\omega/\gamma_{a_\mathrm{d}}=0.1$ (blue); $0.05$ (orange). Error bars indicate stochastic uncertainties. 
    }
    \label{fig:epsilonGammaDark}
\end{figure}

Any consideration to further improve the detection efficiency of our measurement scheme has to start by recognizing the hard upper limit of $81 \%$ for a resonant Gaussian single-photon pulse achieved by perfect projective measurements stroboscopically repeated at the optimal rate as discussed in Sec.~\ref{sec:Proj}.
One standard approach to improve the efficiency of a nonabsorbing detector is to cascade several measurement units. While the JPD measurement process is indeed nonabsorbing, the stroboscopic measurements scatter the photons in a multitude of modes and disturb the shape of the quantum pulse so severely, that any downstream cavity will struggle to absorb the photon and thus cannot accrue further detections.

Instead, we consider a setup, where a single photon pulse is preamplified before being fed into our JPD detector. If two (slightly-delayed) independent photons were produced from a single incoming photon, the detection limit could go up to $96 \% = 1-(1-81\%)^2$.
The preamplification scenario is particularly germane, as the very same two-cavity JPD circuit constituting our detector can act as a photon-number amplifier: operated at a biasing condition 
 $2e V_\mathrm{dc}/\hbar = n \omega_b - \omega_a$ ($n \in \mathbb{N}$) each tunneling Cooper pair creates $n$ excitations into cavity $b$ by annihilating one excitation from cavity $a$ \cite{leppakangas_multiplying_2018, albert_microwave_2024, danner_amplification_2025}. The Hamiltonian~\eqref{eq:full_H}
becomes in the appropriate frame and RWA  
\begin{equation}
      \label{eq::1stage_Hamiltonian_RWA}
      \hat{H}_{J,\mathrm{RWA}} 
      \approx \frac{E_J^* \alpha_0 \beta_0^n}{2 n!} (\ann \bdag[n] + \adag \bnn^n)
\end{equation}
where higher-order corrections similar to the Bessel-functions in \eqref{eq:RWA_H} and not relevant to our discussion will appear for larger zero-point fluctuations $\alpha_0, \beta_0$. 
When the Josephson coupling of the multiplier is properly matched to in- and output coupling rates \cite{leppakangas_multiplying_2018}, an incoming resonant photon will with near certainty enter the cavity $a$ and be multiplied by a tunneling Cooper pair to $n$ photons in cavity $b$, which will subsequently leak out of this cavity. The modes of this output and their occupation can be further adjusted by modifying the ratio of input and output coupling, $\gamma_a/\gamma_b$, cf.~Ref.~\cite{danner_amplification_2025} and App.~\ref{app:multiplier}.

As a preliminary to studying the full cascaded set-up shown in Fig.~\ref{fig:MultiCascSpec} 
we again turn to our toy-model for the detection process, and consider the preamplifier's output impinging on a cavity, which is stroboscopically subjected to perfect projective measurements. 
We chose a driving strength $E_{J_m}$ for the multiplier that guarantees near-perfect multiplication \cite{leppakangas_multiplying_2018, danner_amplification_2025} and reduce the frequency broadening of the incoming pulse to retain good absorption, as the multiplication device leads to a slight broadening, see App.~\ref{app:multiplier}.

For a multiplication factor of $n=2$ 
a detection probability of $\eta=95 \%$ is reached for the optimal measurement rate (c.f.~black dots in Fig.~\ref{fig:project_opt}), which stays very slightly below the $96 \%$ predicted for two independent photons. 
Similarly, for $n=3$ we find $98\%$ ($99\%$ respectively). 

Results for the full cascaded device shown in Fig.~\ref{fig:MultiCascSpec} for an $n=2$ multiplier (orange line) are compared in Fig.~\ref{fig:epsilonGammaDark} to results without preamplification (blue line). Here, for fixed device parameters and weight function the threshold is increased from left to right, yielding increased detection efficiency, but also rising dark rates. The improvement by preamplification is substantial: for instance for a dark rate of $\gamma_\mathrm{dark}=0.116 \cdot 10^{-3} \gamma_{a_\mathrm{d}}$ the detection efficiency increases from $\eta_1=69.8 \%$ without multiplier to $\eta_2=88.5 \%$ with preamplification. 

Note, however, that we show preamplifier results for a photon rate reduced by a factor of two ($6.25 \cdot 10^{-3} \gamma_{a_\mathrm{d}}$ necessitated by the reduced frequency broadening of the pulse, see App.~\ref{app:multiplier}) and that dark counts stemming from spontaneous photon creation in $a_\mathrm{d}$ (see Sec.~\ref{sec:JPDStrobo} and App.~\ref{app:RWA}) remain a problem. 

Due to a large computational effort to simulate the full system, simulations are limited to small multiplication factor $n \lesssim 2$ and an exhaustive investigation of the large parameter space and thorough optimization are beyond the scope of this study. This is particularly regrettable as the naive prediction for two-photon efficiency, $\eta^\textrm{naive}_2 = 1-(1-\eta_1)^2$ (blue dashed line in Fig.~\ref{fig:epsilonGammaDark}) holds surprisingly well. This suggests, that the same argument may be used to calculate n-photon efficiencies for larger multiplication factor $n$ and further improvements towards near-deterministic single-photon detection may be expected. 

\section{Conclusion and Outlook}
We presented a novel scheme to detect a single itinerant microwave photon by stroboscopic projective measurement of a cavity, on which it impinges. We showed that a near-projective measurement can be realized by exploiting the nonlinearity of a Josephson junction, which couples the cavity receiving the photon to a second one within a dc-voltage biased circuit.
A driving, which depends on whether the photon has been absorbed by the cavity or not, is induced by the junction on the second cavity, whose output, monitored by homodyne detection, reveals the photon detection.

A toy-model of perfect instantaneous projective measurement was used to find the stroboscopic rate for which repeated measurements maximize the chance of detection, while not curtailing absorption too severely by the quantum Zeno effect.

These results informed the full measurement scheme, where a strong coupling is switched on for short times interrupted by long uncoupled intervals. Choosing a suitable weight function and threshold for the output signal, photon detection efficiencies and dark count rates could be calculated. The detection scheme was amended by using a similar Josephson-circuit as a preamplifier. 

Using realistic parameters for the experimental devices, we found, for instance for a dark count rate of $\gamma_\mathrm{dark}=0.116 \cdot 10^{-3} \gamma_{a_\mathrm{d}}$, a photon detection efficiency of $\eta_1=69.8 \%$, rising to $\eta_2=88.5 \%$ using a preamplifier with a multiplication factor of $n=2$. In fact, the experimental devices described in this proposal have already been used in the multiplier configuration for the amplification of a (classical) coherent continuous wave input in \cite{jebari_near-quantum-limited_2018, albert_microwave_2024}. Besides the increased complexity of integrating a photon source and extending to the cascaded scenario, the sufficient suppression of unwanted off-resonant processes may be experimentally most challenging. 
The flexibility of the Josephson-photonics devices will however allow us to easily increase the multiplication factor for preamplification.
In fact, this can in principle be done in-situ by a simple change of the dc voltage of the preamplifier device. 
In a similar fashion, different resonance conditions can be chosen by the dc voltage of the detector device to activate alternative effective coupling terms between the cavities. Different measurement strategies for these can then be devised and investigated. 
Progress in the signal analysis can also be made by further optimizing weighting functions and thresholding rules. 
The versatile platform of Josephson-photonics devices holds high promises and the potential to play a substantial role for the number-resolved detection of itinerant microwave photons.

\section{Acknowledgements}

MH acknowledges financial support from the Natural Sciences and Engineering Research Council of Canada (NSERC) through grant RGPIN-2025-06130. 
JA gratefully acknowledges the support of the Deutsche Forschungsgemeinschaft (DFG, German Research Foundation) through AN336/17-1 and AN336/18-1 and the Bundesministerium für Bildung und Forschung BMBF) through QSolid.
HZ acknowledges support by the state of Baden-Württemberg through bwHPC
and the German Research Foundation (DFG) through grant no INST 40/575-1 FUGG (JUSTUS 2 cluster).

\appendix

\section{Reverse exponential pulse}
\label{app:rev_exp_pulse}

\begin{figure}[t]
    \centering
    \includegraphics[width=\columnwidth]{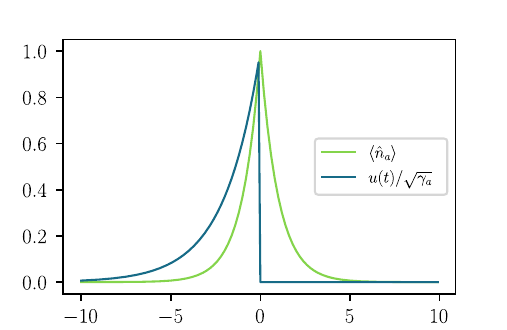}
    \caption{Occupation of a cavity (green) with loss rate $\gamma_a$ on which a single photon in a reverse exponential pulse (blue) impinges. Its time constant matches the life time of the cavity. 
    At $t=0$, the occupation of the cavity reaches unity. 
    }
    \label{fig:revExp}
\end{figure}

For successful photon detection in a projective measurement scheme it is essential, that the interrogated cavity is driven by the impinging pulse to as high a probability of occupation as possible. Indeed, it is well known, that there exists a specific pulse shape, a reverse exponential pulse $u(t)= \sqrt{\gamma_a} \, \theta(-t) \,  \ee^{\gamma_at/2}$, that achieves perfect occupation of the Fock-state $\ket{1}$ at $t=0$, see Fig.~\ref{fig:revExp} showing pulse shape, perfect reflectionless absorption and the subsequent decay. Intuitively, this observation can be understood as a consequence of time-reversal symmetry: the considered scenario being the time-reverse of the natural decay of cavity with constant loss rate $\gamma_a$ initialized in $\ket{1}$ to the environment.

If cavity $a$ is measured projectively once, exactly at $t=t_0$, a photon is detected for sure, $\eta=100 \%$, as indicated by the upper border of the red shaded area in Fig.~\ref{fig:project_opt} for $\gamma_\mathrm{m}\rightarrow 0$. Increasing the measurement rate and performing several measurements during the pulse disturbs the process of perfect absorption, and the maximal total detection probability decreases. More relevant for a detection scheme is the average over all possible positions of the stroboscopic measurements, for which we found in Fig.~\ref{fig:project_opt} (red line) rather low detection efficiencies for the time-reversed exponential, partly explained by its rather wide spectral width which, by construction, cannot be much smaller than the cavity linewidth. 

\section{Weight function }\label{app:weight}

\begin{figure}[b]
    \centering
    \includegraphics[width=\columnwidth]{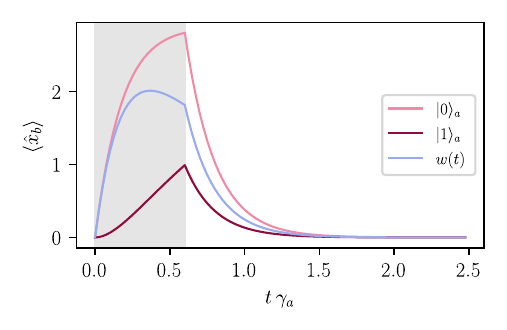}
    \caption{Expectation values $\avg{\hat{x}_b}$ for initial state $\ket{0_a 0_b}$ and $\ket{1_a 0_b}$ according to a Lindblad master equation. 
    The gray area indicates the duration of the measurement ($E_J\neq0$).
    The weight function $w(t)=\avg{\hat{x}_b}_{0}-\avg{\hat{x}_b}_{1}$ is calculated as the difference of the two Lindblad solutions. 
    Parameters as in Fig.~\ref{fig:threshold}.}
    \label{fig:weight}
\end{figure}

The Measurement outcomes $O_j$ are calculated by a weighted integral over $J_{x_b}$ as shown in Eq.~\eqref{eq:Oi}. 
In the following, the origin of the weight function is explained in more detail. 
Conceptually, it is clear that at the very beginning of the measurement, when $E_J$ is switched on, there is no information in $J_{x_b}$ since in both cases (photon or no photon in $a$), cavity $b$ starts out in the same state. 
Therefore, the start of the measurement should be weighted little and the weight should increase while $b$ is approaching its steady state, which is dependent on the state in $a$. 

On the other hand, cavity $a$ is lossy as well. 
Therefore, the photon in $a$ might get lost while $E_J$ is non-zero which warrants a decrease of the weight towards the end of $E_J\neq 0$. When $E_J=0$ again, there is still information in the amplitude of $b$ as it needs time to relax back to the vacuum state. During this, the weight function should be non-zero but decay on a scale of $\gamma_b$.  

To fulfill these specifications, we calculate the weight function as the difference between the expectation values $\langle\hat{x}_b\rangle$ for a Lindblad solution, where the initial state was either $\ket{0_a 0_b}$ or $\ket{1_a 0_b}$, see Fig.~\ref{fig:weight}. 
Starting from the initial state $\ket{0_a 0_b}$, cavity $b$ approaches its steady state of $\ket{\beta}$ very fast, on a time-scale defined by $\gamma_b$, while for $\ket{1_a 0_b}$, this state is reached on the much slower time scale given by the loss of the photon from $a$ with  rate $\gamma_a$. 

Further optimization of the evaluation of the measurement signal, specifically the exact form of the weight function, is left for future work. 

\section{Validity of the rotating-wave approximation}\label{app:RWA}

\begin{figure}[b]
    \centering
    \includegraphics[width=\columnwidth]{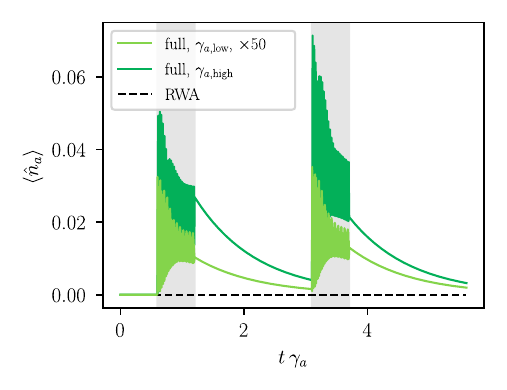}
    \caption{Occupation $\avg{\hat{n}_a}$ according to a Lindblad master equation for two measurements without input pulse. 
    For the RWA Hamiltonian \eqref{eq:RWA_H}, cavity $a$ stays empty since there is no mechanism in the system that could excite the cavity. 
    For low bandwidth (light green), the occupation  $\avg{\hat{n}_a}$ stemming from the full Hamiltonian \eqref{eq:full_H} closely matches the RWA results with a maximum occupancy of $\avg{\hat{n}_a}_\mathrm{max}= 7 \cdot 10^{-4}$. 
    For a higher bandwidth (dark green), clear deviations from the RWA are visible with 
    $\avg{\hat{n}_a}_\mathrm{max}=0.07$.
    This would lead to a high dark rate in the investigated measurement scheme.\newline
    [Parameters: 
     $\gamma_{a, \mathrm{low}} = 10^{-4} \,\omega_a$, $\gamma_{a, \mathrm{high}} = 10^{-3}\omega_a$, $\omega_b = 0.76\omega_a$, other parameters see Fig.~\ref{fig:threshold}.] 
    }
    \label{fig:RWAValid}
\end{figure}

As discussed in the main text, the preferred regime for detecting photons with the proposed scheme is a large ratio of $\gamma_b/\gamma_a$, so that the cavity $b$ can quickly adapt to changes of $a$ and therefore approaches an instantaneous measurement.  
Furthermore, during a measurement the change of amplitudes of $\hat{x}_b$ for $\ket{\psi}_a=\ket{0}$ and $\ket{\psi}_a=\ket{1}$ must be so large that the two states can be distinguished even with the added noise from the homodyne measurement. 
To fulfill these two conditions simultaneously a large $E_J$ is needed. This, however, puts in doubt the validity of replacing the full (lab-frame) Hamiltonian \eqref{eq:full_H} of the main text, by the RWA-Hamiltonian \eqref{eq:RWA_H}.   
The validity of the rotating wave approximation is gauged by comparing the scale of $E_J$ with all frequencies, $\omega_a,\omega_b=\omega_\mathrm{dc}$ and all  their sums and differences, such as $|\omega_a-\omega_\mathrm{dc}|$.

In the rotating frame, the Hamiltonian contains terms of the form $(\adag)^k \bnn^l$ or $\ann^k \bnn^l$ for $l,k \in \mathbb{N}_0$ with prefactor $E_J^*/2 \alpha_0^k \beta_0^l$ (and their conjugated complex counterparts. 
These terms rotate with the frequency $\omega_\mathrm{rot}=k \omega_{a,\mathrm{rot}} - l\omega_{b,\mathrm{rot}} \pm \omega_\mathrm{dc}$ and $\omega_\mathrm{rot}=-k \omega_{a,\mathrm{rot}} - l\omega_{b,\mathrm{rot}} \pm \omega_\mathrm{dc}$, respectively, and are assumed negligible within RWA when $E_J^*/2 \alpha_0^k \beta_0^l \ll \omega_\mathrm{rot}$. 

To explain qualitatively which effects the breakdown of the RWA will have on the detection scheme, we exemplarily consider the term  proportional to $\ann$, namely an extra term in the (rotating frame but non-RWA) Hamiltonian, 
\begin{align}
     \ham_{\ann}=&
        - \ii  \frac{E_J^* \alpha_0}{2} \vcentcolon \,\left[  \ann \ee^{-(\omega_{a,\mathrm{rot}}+\omega_\mathrm{dc})t}- \adag \ee^{(\omega_{a,\mathrm{rot}}-\omega_\mathrm{dc})t} \right] \nonumber \\
        & \frac{\mathrm{J}_1\left(2\alpha_0 \sqrt{\hat{n}_a} \right)}{ \alpha_0 \hat{n}_a^{1/2}}  \mathrm{J}_0\left(2\beta_0 \sqrt{\hat{n}_b} \right) \, \vcentcolon \ . 
\end{align}
In case this term is non-negligible, there will be spontaneous excitations in cavity $a$ while performing measurements. 
These excitations are not distinguishable from itinerant photons getting absorbed by $a$ in our measurement scheme, and therefore these excitations will yield additional incorrect photon counts and increase the dark count rate.

The effects of this specific process along with all other terms and associated processes included in the full time-dependent Hamiltonian \eqref{eq:full_H} are illustrated by Fig.~\ref{fig:RWAValid}, where we compare scenarios with differing $\gamma_a$,  which effectively sets the bandwidth of detectable photons.  

For a fixed central operating frequency $\omega_a$, a higher $\gamma_a$ implies a larger $E_J$ (as we need to keep $\gamma_a \ll \gamma_b$ and $\beta=E_J^* \beta_0/( \hbar\gamma_b)$ approximately constant)  and, hence, larger non-RWA effects. Indeed, there are significant excitations in cavity $a$ (see the rapidly oscillating dark green line in Fig.~\ref{fig:RWAValid}, and a maximum occupation of $\avg{\hat{n}_a}_\mathrm{max}= 0.07$ is reached. Moreover, after the first measurement has concluded, the occupation does not fully decay back to zero before the second measurement occurs further increasing the probability of subsequent dark counts. 
Decreasing $\gamma_a$ at the cost of lowering the bandwidth strongly suppresses the effects of the rotating terms, and the result follows RWA more closely (light green line). The maximum occupation in this case stays below $\avg{\hat{n}_a}_\mathrm{max}= 7 \cdot 10^{-4}$. 
This value can be used as a rough estimate for the probability to detect a photon in a single measurement, without input photon (dark count). 
In consequence, we expect an additional dark rate,  that is not captured by the RWA calculations presented in the main text, of $\gamma_\mathrm{dark,rot}=\avg{\hat{n}_a}_\mathrm{max} \gamma_\mathrm{m}=2.8  \cdot 10^{-4} \gamma_a$, which is of similar order of magnitude as the dark count rate obtained there from the full simulations of the measurement process within RWA. 

\section{Multiplier}
\label{app:multiplier}

\begin{figure}[t]
    \centering
    \includegraphics[width=\columnwidth]{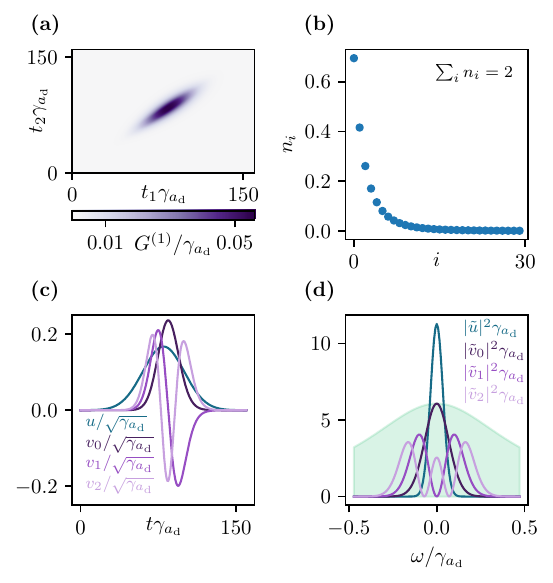}
    \caption{(a) Autocorrelation function $G^{(1)}(t_1,t_2)$ of the output of the multiplier for multiplication factor $n=2$. 
    (b) Occupation $n_i$ of the modes found from the diagonalization of $G^{(1)}(t_1,t_2)$. The total occupation across all modes is $2$. 
    (c) Temporal mode shapes of input mode $u(t)$ and the three most occupied modes $v_i(t), \, i=0,1,2$. 
    (d) Frequency spread of input mode $\tilde u(\omega)$ and the three most occupied modes $ \tilde v_i(\omega), \, i=0,1,2$ in comparison to the resonance curve of cavity $\ann_\mathrm{d}$ (turquoise shaded area). 
    [Parameters: $n=2$, $\gamma_{a_\mathrm{m}}/\gamma_{a_\mathrm{d}}=0.5$,$\gamma_{b_\mathrm{m}}/\gamma_{a_\mathrm{d}}=0.5$, $\alpha_{0_\mathrm{m}}=0$, $\beta_{0_\mathrm{m}}=0$, $E_{J_\mathrm{m}}^* \beta_{0_\mathrm{m}}^2 \alpha_{0_\mathrm{m}}/(\hbar \gamma_{a_\mathrm{d}})=1$. 
    For the detection probability in Sec.~ \ref{sec:MultiCasc} of the main text for $n=3$: $E_{J_\mathrm{m}}^* \beta_{0_\mathrm{m}}^3 \alpha_{0_\mathrm{m}}/(\hbar \gamma_{a_\mathrm{d}})=\sqrt{9/2}$.]
    }
    \label{fig:MultiModes}
\end{figure}

A preamplification scheme, dubbed as the inelastic Cooper-pair tunneling photon multiplier \cite{leppakangas_multiplying_2018, albert_microwave_2024}, can be prepended to our detection device. 
It is a similar Josephson-photonics device with two microwave modes, where the resonance condition is set to $\omega_{\mathrm{dc}_\mathrm{m}} + \omega_{\mathrm{a}_\mathrm{m}}\approx n \, \omega_{\mathrm{b}_\mathrm{m}}$. 
Then, an impinging microwave photon is absorbed by cavity $a_\mathrm{m}$ and multiplied to $n$ photons in cavity $b_\mathrm{m}$ by the energy of one inelastically tunneling Cooper pair. 
An impedance-matching condition fixes the driving strength $E_{J_\mathrm{m}}^* = \hbar \sqrt{\gamma_{a_\mathrm{m}}\gamma_{b_\mathrm{m}}} \sqrt{n\,n!}/(\alpha_{0_\mathrm{m}}\beta_{0_\mathrm{m}}^n)$, where perfect conversion ($n$ photons are emitted from cavity $b_\mathrm{m}$ into an output transmission line) can be achieved for a continuous, coherent-state input. 
For incoming temporal pulses that are sufficiently resonant with cavity $a_\mathrm{m}$ (compare Fig.~\ref{fig:MultiModes}), this driving strength still ensures near-perfect conversion. 
In general, the output of the preamplifier is multimode due to the nonlinear multiplication process. 
A single incoming photon with frequency $\omega_\mathrm{in}=\omega_{a_\mathrm{m}}\pm \sigma_\omega $ is multiplied to $n$ outgoing photons that are spectrally broadened by the inverse lifetime $\gamma_{b,\mathrm{m}}$ of cavity $b_m$, because the resonance condition $\omega_{\mathrm{dc}_\mathrm{m}} + \omega_{\mathrm{a}_\mathrm{m}} \pm \sigma_\omega\approx n \, (\omega_{\mathrm{b}_\mathrm{m}}\pm\gamma_{b_\mathrm{m}})$ only fixes the frequency of the sum of the $n$ outgoing photons. 
By tuning the remaining parameter, $\gamma_{a_\mathrm{m}}/\gamma_{b_\mathrm{m}}$, the precise output mode structure can be modified. A spectral decomposition of the two-time coherence function $G^{(1)}(t_1, t_2) = \avg{\bdag_{\mathrm{m}, \mathrm{out}}(t_2) \bnn_{\mathrm{m}, \mathrm{out}}(t_1)}$ of Fig.~\ref{fig:MultiModes}(a) yields the eigenmode occupations and decomposition, c.f.~Fig.~\ref{fig:MultiModes}(b, c) of the emitted radiation into the output transmission line \cite{kiilerich_input-output_2019,kiilerich_quantum_2020}. 
If the photon multiplier itself is used as a photon detector, a regime is favored where a single mode of the output is highly occupied \cite{danner_amplification_2025}. When used as a preamplifier, we turn to a regime of multiple output modes, c.f.~Fig.~\ref{fig:MultiModes}(b). Specifically, these modes are spectrally broader than the original input pulse, Fig.~\ref{fig:MultiModes}(d). If a preamplification step is used, the bandwidth of both devices has to be carefully designed in order to avoid off-resonant reflections of the preamplified signal at the detector. Here, we choose $\gamma_{a_\mathrm{d}}=2\gamma_{a_\mathrm{m}}$, such that most of the preamplified signal is still sufficiently resonant with the detector cavity $a_\mathrm{d}$.

\section*{References}
\bibliographystyle{ieeetr}
\bibliography{references_arXiv}
\end{document}